\documentclass[fleqn,12pt]{wlscirep}

\pdfoutput=1
\usepackage{amssymb}
\usepackage{amsmath}
\usepackage{graphicx,booktabs,array}
\usepackage{url}
\usepackage{xcolor}
\usepackage{setspace}
\newcommand\farcs{\hbox{$.\!\!^{\prime\prime}$}}
\newcommand\farcd{\hbox{$.\!\!^{\circ}$}}
\newcommand{\rmn}[1]{{\mathrm{#1}}}

\title{A peculiar hard X-ray counterpart of a Galactic fast radio burst} 

\author[1]{A.~Ridnaia}
\author[1]{D.~Svinkin}
\author[1]{D.~Frederiks}
\author[1]{A.~Bykov}
\author[2,3]{S.~Popov}
\author[1]{R.~Aptekar}
\author[1]{S.~Golenetskii}
\author[1]{A.~Lysenko}
\author[1]{A.~Tsvetkova}
\author[1]{M.~Ulanov}
\author[4]{T.~Cline}

\affil[1]{Ioffe Institute, 26 Politekhnicheskaya, St Petersburg, 194021, Russia}
\affil[2]{Sternberg Astronomical Institute, Lomonosov Moscow State University, Moscow, 119234, Russia}
\affil[3]{Higher School of Economics, Department of Physics, Moscow, 101000, Russia}
\affil[4]{NASA Goddard Space Flight Center, Greenbelt, Maryland, USA (retired) }

\begin{abstract}
\textbf{Fast radio bursts are bright, millisecond-scale radio flashes of yet unknown physical origin~\cite{Cordes2019}. 
	Recently, their extragalactic nature has been demonstrated~\cite{Ravi2019, Marcote2020}, 
	and an increasing number of the sources have been found to repeat~\cite{CHIMErep}. 
	Young, highly magnetized, isolated neutron stars - magnetars - have been suggested 
	as the most promising candidates for fast radio burst progenitors owing to their 
	energetics and high X-ray flaring activity~\cite{2010vaoa.conf..129P, 2017ApJ...843L..26B}. 
	Here we report the detection with the Konus-\textit{Wind} of a hard X-ray event of April 28, 2020, temporarily coincident with a bright, 
	two-peak radio burst~\cite{NatCHIME, NatSTARE} from the Galactic magnetar SGR~1935+2154 with properties remarkably similar to those of fast radio bursts.
	We show that two peaks of the double-peaked X-ray burst coincide in time with the radio peaks, 
	confirming that the X-ray and radio emission most likely have a common origin.
	Thus, this is the first simultaneous detection of a fast radio burst from a Galactic magnetar and its high-energy counterpart.
	The total energy emitted in X-rays in this burst is typical of bright short magnetar bursts, but an unusual hardness of 
	its energy spectrum strongly distinguish the April 28 event among multiple `ordinary' flares detected from SGR~1935+2154 previously.
	This, and a recent non-detection~\cite{NatSTARE, Kirsten2020, FASTUL} of radio emission from about one hundred typical soft bursts from SGR 1935+2154 
	 favors the idea that bright, FRB-like magnetar signals are associated with rare, hard-spectrum X-ray bursts, which implied rate 
	 ($\sim$~0.04~yr$^{-1}$~magnetar$^{-1}$) appears consistent with the rate estimate\cite{NatCHIME} of SGR 1935+2154-like radio bursts (0.007~--~0.04~yr$^{-1}$~magnetar$^{-1}$).}

\end{abstract}

\begin{document}

\flushbottom
\maketitle	
\thispagestyle{empty}

\section*{MAIN TEXT}

Magnetars~\cite{2015SSRv..191..315M,2017ARAA..55..261K} are rare, young, 
isolated neutron stars with strong magnetic field~\cite{1992ApJ...392L...9D} ($B \sim 10^{14}-10^{15}$~G).
Soft Gamma-ray Repeaters (SGRs), discovered in 1979\cite{1979Natur.282..587M,1979SvAL....5..343M} 
through the detection of repeating short bursts in the hard X-ray/soft $\gamma$-ray range, 
were later associated with the magnetar activities~\cite{1998Natur.393..235K}.  
During an active phase, which may last from several days to a year or more, 
and is followed by a long quiescent period, SGRs sporadically emit short (fraction of a second) 
bursts of hard X-rays, with photon energies below a hundred keV and implied peak X-ray luminosities 
of $L_X \sim 10^{38}-10^{42}$~erg/s. 
Only a dozen burst-emitting magnetars are known so far\cite{Olausen2014} and, up to now, 
no counterpart to such bursts is identified at other wavelengths.

SGR~1935+2154 is a new member of the magnetar family, discovered on 2014 July 5 through a series of short bursts~\cite{Stamatikos2014, Lien2014}. 
Extensive follow-up observations carried out during the first eight months since the source discovery 
with \textit{Swift}/X-ray Telescope, \textit{Chandra} and \textit{XMM-Newton} X-ray observatories confirmed 
that SGR~1935+2154 was indeed a magnetar with a spin period $P \sim  3.24$~s and $\dot P = 1.43 \times 10^{-11}$~s~s$^{-1}$, 
implying a surface dipole magnetic field value of $B \sim 2.2 \times 10^{14}$~G well within the typical range of magnetars~\cite{Israel2016}. 
The source position lies very close to the geometric center of the Galactic supernova remnant (SNR) G57.2+0.8~\cite{Gaensler2014}, 
whose distance estimates range from 6.6~kpc\cite{Zhou2020} up to 12.5~kpc~\cite{Kothes2018}. 
Accounting for the uncertainty in distance, we adopt 10~kpc in this paper. 
Since its discovery SGR~1935+2154 has been one of the most burst-prolific magnetars.
During the recent activation, started in April 2020, it exhibited multiple bright, short 
X-ray bursts, culminating, on April 27, in a burst `forest' made up of rapid sequences of multiple 
flares during which the count rate never returns to baseline~\cite{GCN27663, GCN27665, GCN27667}.

The next day, at $\approx$14:34:33~UT, a bright, millisecond-scale radio burst from the direction of SGR~1935+2154 was detected~\cite{NatCHIME} by CHIME/FRB backend in the 400-800 MHz band.
The CHIME/FRB light curve\cite{NatCHIME} had a two-peak structure with two components, $\sim0.6$~ms and  $\sim0.3$~ms wide, separated by $\sim29$~ms.
Independently, the second pulse was detected~\cite{NatSTARE} by STARE2 in the 1.4~GHz band, with the intrinsic duration of $\sim0.6$~ms.
An inferred fluence of the burst, which properties are remarkably similar to those of fast radio bursts (FRBs), 
was estimated to $\sim 700$~kJy$\cdot$ms by CHIME/FRB\cite{NatCHIME}, and to $\sim$1.5 MJy$\cdot$ms by STARE2\cite{NatSTARE}. Simultaneously, the hard X-ray burst was observed with INTEGRAL~\cite{Mereghetti2020}, 
AGILE~\cite{NatAstrAGILE}, Konus-\textit{Wind}~\cite{GCNKWFRB}(KW), and \textit{Insight}-HXMT~\cite{NatInsight}. 
The burst triggered KW at $T_0$= 14:34:24.447 UT (geocentric time, see Methods). 

The burst triggered KW at $T_0$= 14:34:24.447 UT (geocentric time, see Methods). 
The KW light curve (Figure~1) shows a gradually rising, double-peaked pulse. 
Using a Bayesian block decomposition of the KW count rate in the 18-320~keV band (see Methods), 
we found the emission start and stop times to be $T_0$ - 0.220~s and $T_0$ + 0.244~s, respectively.
The total duration of the burst is 0.464~s. 
The brightest part of the event spans the interval between $T_0$-0.060~s and $T_0$+0.036~s 
and shows two prominent peaks, 0.016~s and 0.032~s wide, which overlap in time with the two radio pulses.  
Although the hard X-ray peaks observed by KW are wider than the sub-millisecond radio peaks, 
the position of their maxima in the KW 18--80~keV light curve coincide with the de-dispersed geocentric 
arrival times of the two radio pulses\cite{NatCHIME,NatSTARE} to within $\pm$2 ms.
Thus, we conclude that the KW event and the radio burst originate from the same source 
and we witness the first simultaneous detection of an FRB-like burst from a Galactic magnetar and its hard X-ray counterpart. 

The photon energies measured by KW during this event extend to beyond 250~keV.
Our spectral analysis (Methods) shows, that the energy spectrum in the 20--500~keV band is well described 
by two spectral models: an exponentially-cutoff power law function (CPL) and a sum of two blackbody functions (2BB). 
For the time-averaged spectrum (from $T_0$ to $T_0+0.256$~s), the CPL low-energy photon index $\alpha$ is $-0.72_{-0.46}^{+0.47}$ 
and the peak energy in $\nu F_\nu$ spectrum $E_p$ is $85_{-10}^{+15}$~keV (all spectral parameter errors are given at the 90\% confidence level). 
To examine the possible presence of a hard power-law component, we fit this spectrum by a joint smoothly broken power-law function 
(the Band GRB function, see Methods). The fit with this model results in nearly the same low-energy photon index and $E_p$, 
with only an upper limit on the high-energy photon index ($\beta <-2.7$).
From the fit with two blackbody functions, the cold and hot blackbody temperatures are estimated to $11_{-4}^{+3}$~keV and $31_{-7}^{+12}$~keV, respectively. 

Derived from our analysis, the CPL photon index $\alpha$ has relatively large uncertainty. 
Despite several orders of magnitude huge confidence bounds of the best-fit CPL spectral model extrapolated to the radio band, 
our data are inconsistent with a single power law spectrum of the event extending from the radio to hard X-rays (Figure~2). 
This result is not surprising, given the signatures of narrow bandwidth fluctuations observed in radio spectrum of the
event~\cite{NatCHIME,NatSTARE} and the findings of recent studies that show that FRBs are not always broadband~\cite{Law2017, Gourdji2019,Kumar2020}.

The broad spectra of the two radio burst components detected by CHIME/FRB are strikingly different\cite{NatCHIME}, 
with the first detected primarily at frequencies under 600 MHz and the second above (in agreement with the non-detection of the first burst at 1.4 GHz by STARE2\cite{NatSTARE}). 
Contrary, the X-ray emission detected by KW  exhibits no pronounced spectral variation between the two peaks (Figure~1),
suggesting that spectral behavior of magnetar radio bursts are not directly connected to that of the accompanying hard X-ray bursts. 

The burst total energy fluence $S$, measured in the 20--500~keV band over the total duration of 0.464~s, is $(9.7 \pm 1.1) \times 10^{-7}$~erg/cm$^2$.
The 4~ms peak fluxes in the first and the second X-ray pulses are $(7.5 \pm 1.2) \times 10^{-6}$~erg/cm$^2$/s 
and $(9.1 \pm 1.5) \times 10^{-6}$~erg/cm$^2$/s, respectively.
For the source distance 10~kpc we estimate the burst total energy release in X-rays to $\approx 1.2\times10^{40}$~erg 
and the peak isotropic X-ray luminosity to $L_X \approx 1.1\times10^{41}$~erg/s, both typical of bright, short magnetar flares. 
From the KW detection, and the inferred radio energetics\cite{NatCHIME,NatSTARE} 
($L_r\sim(7\times10^{36} - 4\times10^{38})$~erg/s, $E_r\sim(3\times10^{34} - 2\times10^{35})$~erg),
the ratio of radio to the peak X-ray luminosity $L_r/L_X$ estimates to $\sim (10^{-5} - 10^{-3})$, 
and the ratio of radio to X-ray energy $E_r/E_X$ is $\sim (10^{-6} - 10^{-5})$.
We note, that the implied $L_r/L_X$ in the April 28 flare is of the same order, 
as in the case of FRBs, if we relate them to magnetar giant (and hyper) 
flares~\cite{1979Natur.282..587M, 2005Natur.434.1098H, 2007AstL...33....1F}
with $L_X \sim (10^{44} - 10^{47})$ erg/s, as radio luminosities of FRBs are estimated 
to be in the range $L_r \sim (10^{38} - 10^{42})$ erg/s \cite{Cordes2019}.

 Although energy scales of the SGR~1935+2154 radio burst and extragalactic FRBs are different by several orders of magnitude, the  discovery of the April 28 event makes an exciting breakthrough towards closing the energy gap between Galactic magnetars and cosmological FRBs, providing new support to magnetar FRB models. 
Several distinct mechanisms have been proposed to explain properties of simultaneous radio and X/$\gamma$-ray emission in the variety of magnetar models of FRBs. They can be divided into two broad groups based on the regions where the observed emission is generated: whether it is in the magnetosphere of magnetar or far away at the surrounding medium interacting with a relativistic outflow (see Emission scenarios section in Methods). Detailed modelling of the processes in the frame of the outlined scenarios are certainly needed and the presented observations of SGR~1935+2154 are tightly constraining the models under construction.

In future, with more detections of coincident radio and X-ray bursts, the proximity of
Galactic magnetars would allow to distinguish between different proposed scenarios
of emission, and compare it with the data on FRBs. Unfortunately, with existing all-sky gamma- and X-ray 
monitors even hyperflares of magnetars hardly can be detected from distances $>$100~Mpc \cite{Svinkin2015}. 
May be, a more realistic approach can be related to radio observations 
of nearby galaxies at distances $\lesssim$few~Mpc, as in this case flares 
like the one from SGR~1935+2154 can produce radio flux of $\sim$few~mJy. 
Note, that FRB-like radio transients already have been detected from the Andromeda galaxy (M31) \cite{2013MNRAS.428.2857R}, 
and also there are known candidates for magnetar giant flares from M31\cite{2008ApJ...680..545M} 
and two other galaxies nearby~\cite{2007AstL...33...19F, GCN27595, GCN27596}.

Finally, we would like to emphasize an unusual spectral hardness of the April 28 X-ray burst. 
While the shape of the burst X-ray spectrum and its energy fluence are in line with other magnetar flares, 
the emission in this event extends up to a few times higher energies than typical ($\gtrsim$250~keV vs. $\lesssim$100 keV)
and, accordingly, the derived $E_p\sim85$~keV is the highest among all the bursts observed from SGR~1935+2154 
so far, including the highly-energetic, $\sim 1.7$~s long intermediate flare~\cite{Kozlova2016} (Figure~3).
The probability of the FRB-associated burst to have $E_p$ drawn from the combined sample of 41~\textit{Fermi}-GBM and 21~KW 
SGR~1935+2154 bursts is $\sim2\times10^{-10}$, highlighting its peculiarity with respect to the rest of the population.
This, and a recent non-detection~\cite{NatSTARE, Kirsten2020, FASTUL} of radio emission from about one hundred 
typical, soft-spectrum bursts from the same source, may imply a link between bright, FRB-like magnetar bursts and atypical, harder X-ray bursts.

This idea finds further support from the statistics of hard X-ray flares observed from a wider magnetar sample.
KW operates in interplanetary space since November 1994, with essentially unocculted view of the entire sky.
Among $\sim$250 relatively bright ($S \gtrsim 4\times 10^{-7}$~erg/cm$^2$), short SGR bursts detected by KW from 
six prolific magnetars for more than 25 years, there are only five hard-spectrum events with $E_p$ similar to 
or higher than that of the April 28 burst (see Methods and Extended Data Table~4).
We did not find in literature any radio observations simultaneous to these bursts, 
so none of them can be excluded from candidates to Galactic FRB counterparts. 
The implied rate of such spectrally-hard bursts is $\sim 0.04$~yr$^{-1}$~magnetar$^{-1}$, 
which appears consistent with the rate estimate of SGR 1935+2154-like radio bursts, with $E_r>10^{34}$~erg,  
of (0.007~--~0.04~yr$^{-1}$~magnetar$^{-1}$) derived from CHIME observations\cite{NatCHIME}. 


\newpage

\begin{figure}[hbt!]\label{fig:LC}
	\centering
	\includegraphics[width=0.7\columnwidth]{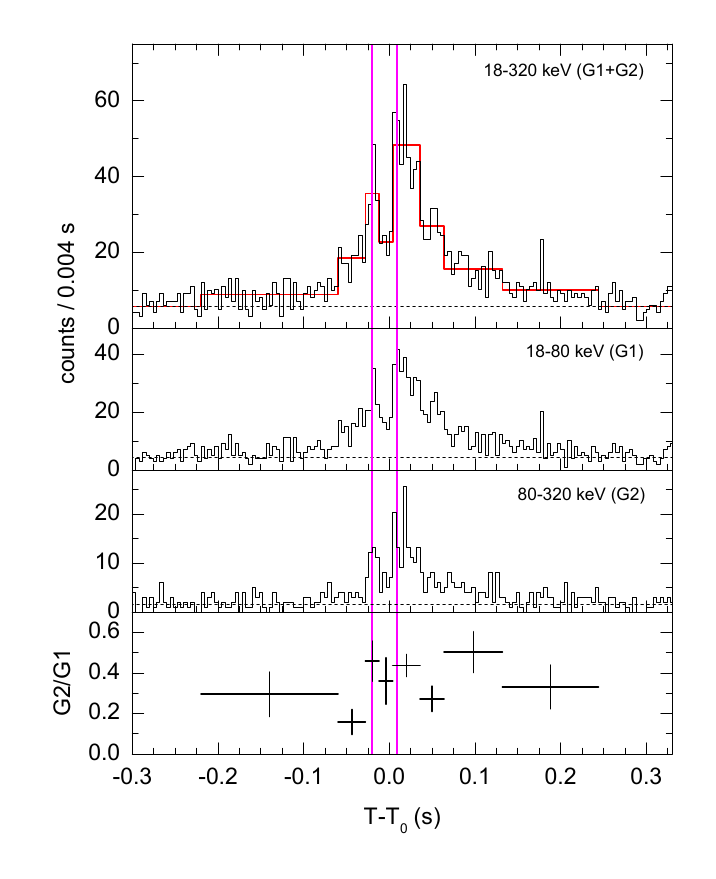}
	\caption*{{\bf Figure 1.} Hard X-ray burst from SGR~1935+2154 of April 28, 2020. 
		The burst time history is shown, as recorded by Konus-\textit{Wind}, in three energy bands: G1~(18-80~keV), G2~(80-320~keV),
		and the combined G1+G2 band~(18-320~keV, top panel). A Bayesian block decomposition of the G1+G2 light curve (see Methods) 
		is shown in the top panel with the thick red line.
		Background count rates are indicated with horizontal dashed lines.  
		In the course of the burst, the emission hardness, illustrated by the evolution of the hardness ratio G2/G1 
		(bottom panel; G2/G1 error bars correspond to 68\% CL) exhibits no pronounced variation, with a hint of the positive hardness-intensity correlation near the main peaks.
		$T_0$ corresponds to 14:34:24.447 UT (geocentric time). 
		The position of the two peak maxima in the KW G1 light curve (14:34:24.427 and 14:34:24.455, respectively) coincide to within $\pm$2 ms with 
		the de-dispersed geocentric arrival times of the two radio components, reported by CHIME/FRB\cite{NatCHIME}: 14:34:24.42650(2) and 14:34:24.45547(2), respectively (shown by vertical lines). 
	}
\end{figure}

\newpage

\begin{figure}[hbt!]\label{fig:SED}
	\centering
	\includegraphics[width=0.99\columnwidth]{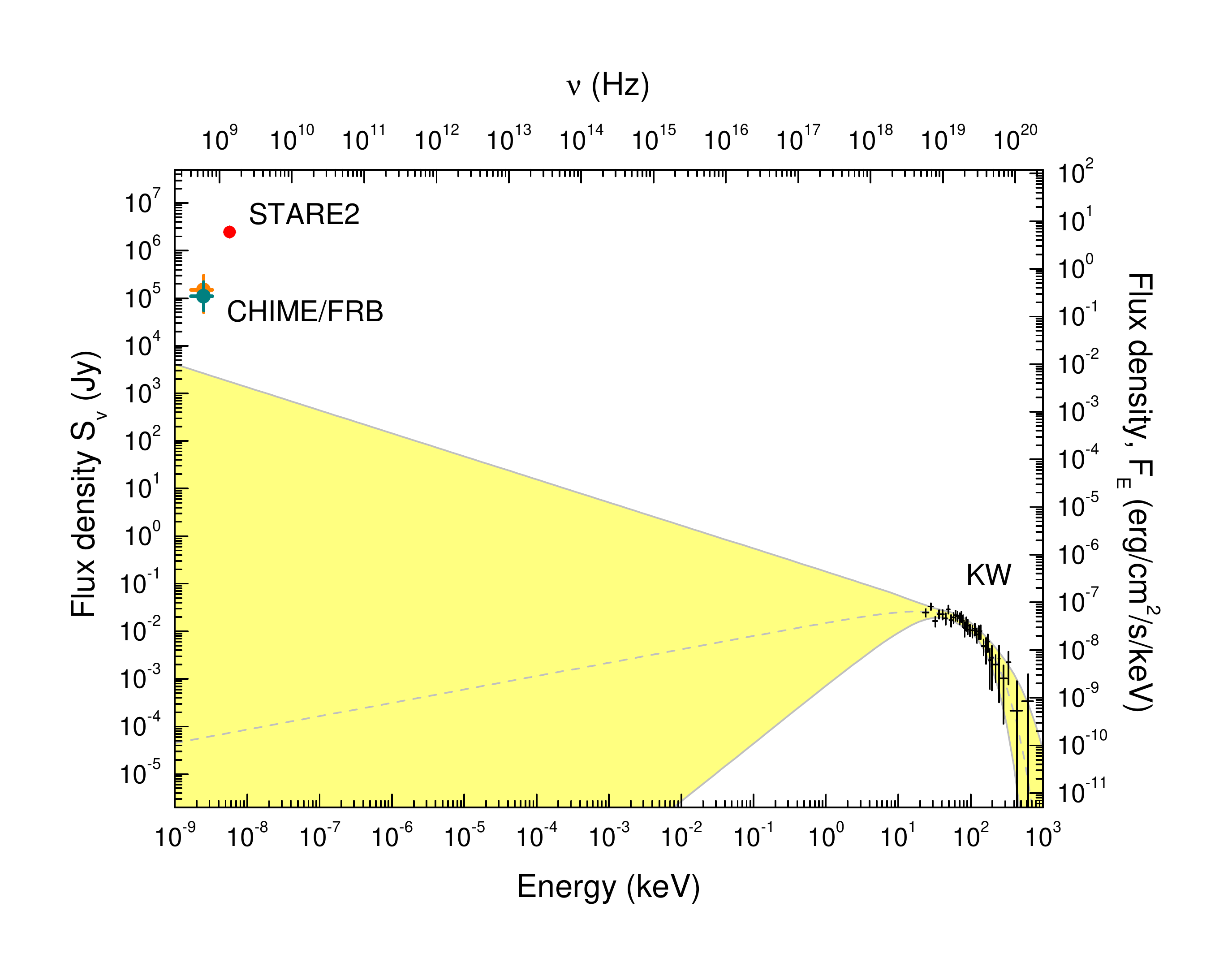}
	\caption*{{\bf Figure 2.} Radio to X-ray spectral energy distribution. 
		The high-energy data points are from Konus-\textit{Wind}; 
		vertical error bars correspond to 1$\sigma$ statistical uncertainties for the data, and horizontal error bars indicate the width of the energy channel.
		The dashed line is the best-fit to the KW data with a CPL function, solid lines correspond to the boundaries of the fit confidence interval at 99.7\% probability. 
		The 400-800~MHz fluxes of the first and the second radio burst components reported by CHIME/FRB~\cite{NatCHIME} are shown, with their reported uncertainties, by green and orange symbols, respectively.
		The 1.4~MHz radio flux (red) is derived from the inferred fluence reported by STARE2~\cite{NatSTARE}.
		Details of the KW spectral analysis are given in Methods and also in Extended Data Figures 1 and 2.
	}
\end{figure}

\newpage

\begin{figure}[hbt!]\label{fig:EpFlux}
	\centering
	\includegraphics[width=0.99\columnwidth]{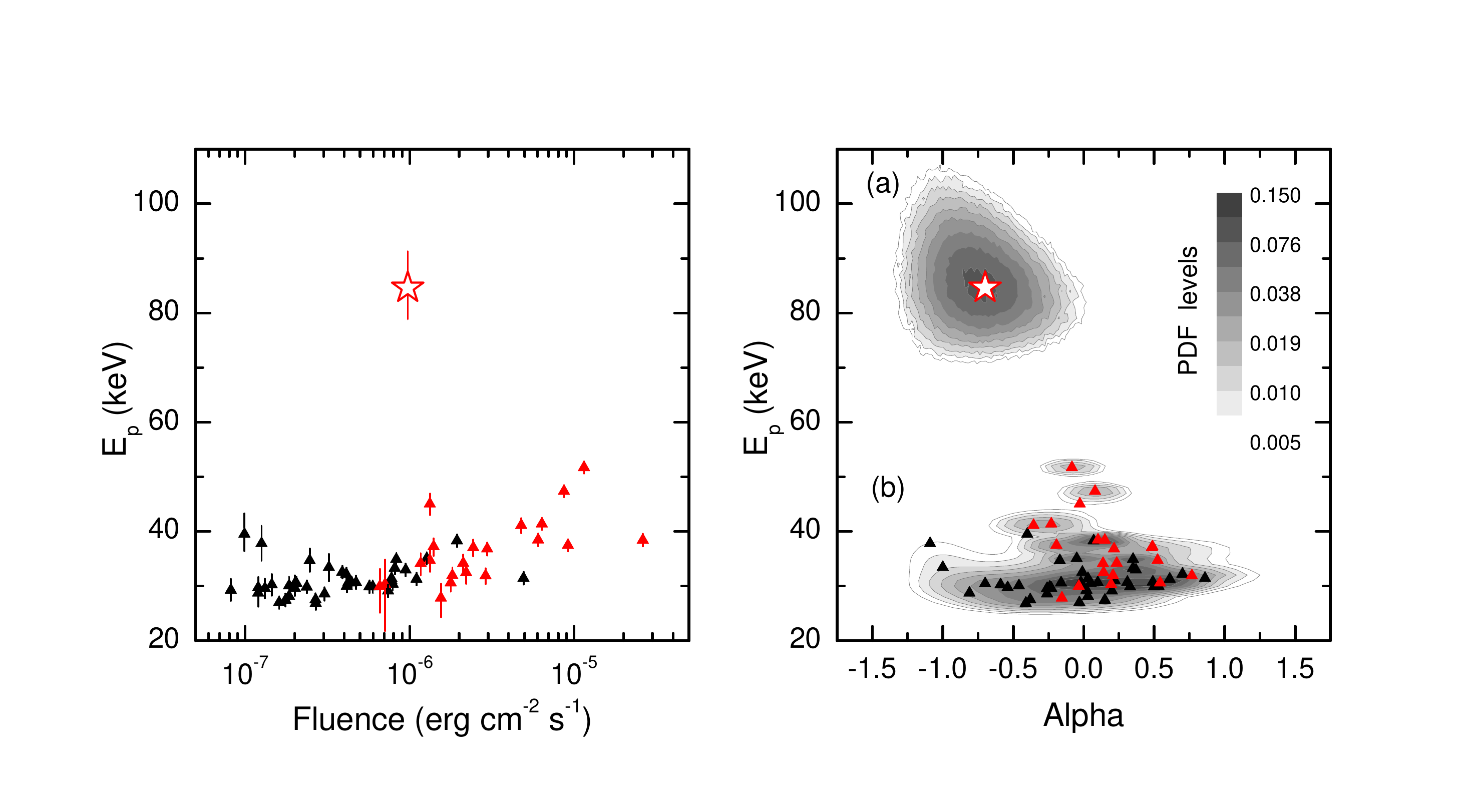}
	\caption*{{\bf Figure 3.} Properties of the April 28 event and other SGR~1935+2154 X-ray bursts.
		Left panel: $E_p$ (the peak energy of $\nu F_\nu$ spectrum) vs. energy fluence.
		21 short X-ray bursts detected by KW in 2015-2020 are shown by red symbols and 
		the red star represents the April 28 burst. Black symbols show 41 \textit{Fermi}-GBM bursts~\cite{Lin2020} detected in 2014-2016. 
		Energy fluences are in the 20--500~keV and 8--200~keV range for the KW and GBM bursts, respectively.
		All error bars are given at the 68\% confidence level. 
		Right panel: Two-dimensional probability density functions (PDFs) for the CPL spectral model parameters are shown in the $\alpha-E_p$ plane
		for (a) the April 28 burst (the red star, see Methods), 
		and (b) the combined sample of 21 KW and 41 GBM bursts (red and black points, respectively). 
		The April~28 burst has the highest $E_p$ ($\sim85$~keV) among all the bursts from SGR~1935+2154, while its energy fluence and photon index $\alpha$ 
		are  within the range of the observed values. 
		The probability of the FRB-associated burst to have $E_p$ drawn from the combined sample of SGR~1935+2154 bursts is $\sim2\times10^{-10}$, 
		highlighting its peculiarity with respect to the rest of the population.
	}
\end{figure}


\newpage

\section*{Methods}
\vspace{0.5cm}
\subsection*{Observations} 
\vspace{0.5cm}
\paragraph*{Konus-Wind}
Konus-Wind\cite{Aptekar_1995SSR} (KW) consists of two identical NaI(Tl) scintillation detectors, 
each with $2\pi$~sr field of view, mounted on opposite faces of the rotationally stabilized \textit{Wind} spacecraft\cite{Harten_1995SSRv}, 
such that one detector~(S1) points towards the south ecliptic pole, thereby observing the south ecliptic hemisphere, 
while the other~(S2) observes the north ecliptic hemisphere.
The April 28, 2020 burst from SGR~1935+2154 triggered detector S2, the incident angle was 47$\farcd$1.

Each KW detector is 5~inches in diameter and 3~inches in height, placed into an aluminum container faced with 
a beryllium entrance window. The crystal scintillator is viewed by a photomultiplier 
tube through a 20~mm thick lead glass, which provides effective detector shielding  
from the spacecraft's background in the soft spectral range. The detector effective area is $\sim 80$--160~cm$^2$, 
depending on the photon energy and incident angle. The energy range of gamma-ray measurements covers 
the incident photon energy interval from 20~keV up to 16~MeV. 

The instrument has two operational modes: waiting and triggered. While in the waiting
mode, the count rates (lightcurve) are recorded in three energy band G1~(18--80~keV), G2~(80--320~keV),
and G3~(320--1300~keV) with 2.944~s time resolution. When the count rate in G2 exceeds a $\approx 9\sigma$ threshold
above the background on one of two fixed time-scales, 1~s or 140~ms, 
the instrument switches into the triggered mode. 

In the triggered mode, light curves are recorded in the same G1, G2, and G3 bands, 
starting from 0.512~s before the trigger time $T_0$ with time resolution varying from 2~ms up to 256~ms.
For the burst of interest, the whole time history is available with the 2~ms resolution. 

Spectral measurements are carried out, starting from the trigger time $T_0$ (no multichannel spectra are available before $T_0$)
in two overlapping energy intervals with nominal boundaries 20--1300~keV (PHA1) and 260~keV--16~MeV (PHA2)
with 64 spectra being recorded for each interval over a 63-channel, pseudo-logarithmic energy scale.
The first four spectra are measured with a fixed accumulation time of 64~ms in order to study short bursts
and they cover about a half of the April 28 burst time history. 

For the presented analysis we use a standard KW dead time (DT) correction procedure for light curves
(with a DT of a few microseconds) and spectra (with a DT of $\sim42$ microseconds).
The burst of interest has a moderate photon flux of $\sim 10^4$~counts~s$^{-1}$,
so no additional correction, which was used, e.g., in the analysis~\cite{Mazets1999AstL} 
of the KW detection of the 1998 August 27 giant flare from SGR~1900+14, is required.

\paragraph*{The burst arrival times in X-ray and radio bands}
The burst triggered KW on 2020-04-28 at $T_0$ = 14:34:24.084~UTC (52464.084~seconds since midnight).
The burst data became available on ground at about 2020-04-28 20:00 UTC.

The \textit{Wind} ephemeris at $T_0$  R.A. = 22$\farcd$502, Dec. = 9$\farcd$547 (J2000), 
and the distance from the Earth center of 1326000.1~km were calculated using 
the \textit{Wind} predicted ephemeris. The ephemeris data and their description are available at 
\url{https://spdf.gsfc.nasa.gov/pub/data/wind/orbit/pre_or/} and  
\url{https://cdaweb.gsfc.nasa.gov/misc/NotesW.html#WI_OR_PRE}, respectively.
The \textit{Wind} clock accuracy is measured with a few day interval, 
the measurement closest to the burst time is available in  
\url{ftps://pwgdata.sci.gsfc.nasa.gov/pub/wind\_clock/wind_20-119-13-21-00.cc_rpt}.
For the burst detection date the estimated clock correction was 2.7~ms (the lag of the \textit{Wind} on-board clock), 
so, the corrected KW trigger time is $T_0=$~14:34:24.087~UTC (52464.087~s).

Using the SGR~$1935+2154$ position\cite{Israel2016} R.A. = 19$^h$34$^m$55$\farcs$5978s, Dec. = +21$^{\circ}$53$^{\prime}$47$\farcs$7864 (J2000.0)
we calculated the Earth-crossing time (the KW trigger time corrected for the \textit{Wind}-Earth propagation time) to be 
$T_{0, \textrm{KW, Earth}}$ = 14:34:24.447~UTC (52464.447~s).

\subsection*{Data analysis}
\vspace{0.5cm}
\paragraph*{Temporal analysis}  
For the temporal analysis we used time histories from $T_{0}-0.512$~s to $T_0+0.512$~s in three energy bands: 
G1 (18--80~keV), G2 (80--320~keV) and G3 (320--1300~keV), with a time resolution of 2~ms .
The total burst duration $T_{100}$, and the $T_{90}$ and $T_{50}$ durations
(the time intervals that contain 5\% to 95\% and 25\% to 75\% of the total burst count fluence, 
respectively~\cite{Kouveliotou_1993ApJ}, see Extended Data Table 1), were calculated in this work using
the light curves in G1 and G2 energy band. Burst start and end times in the each 
band were calculated at the $5 \sigma$ with a method similar to that developed
for BATSE~\cite{Koshut_1996}. The background count rates, estimated using 
the data from $T_0+8.448$~s to $T_0+98.560$~s, are 1056~s$^{-1}$ (G1), 380~s$^{-1}$ (G2), and 173~s$^{-1}$ (G1).

\paragraph*{Spectral analysis}
We analyzed two energy spectra, accumulated from $T_0$ to $T_0+0.064$~s (the peak spectrum) and from $T_0$ to $T_0+0.256$~s (time-integrated spectrum), 
which approximately correspond to the second peak and the entire burst emission after $T_0$, respectively.
The background spectrum was extracted from the interval $T_0+8.448$~s to $T_0+98.560$.

We performed the spectral analysis in XSPEC, version 12.10.1\cite{Arnaud1996}, 
using the following spectral models:
a simple power law (PL), a custom exponential cutoff power-law (CPL) parametrized with peak of $\nu F_{\nu}$ spectrum and energy flux as the model normalization,
and the Band GRB function\cite{Band_1993ApJ}. We also used single blackbody (BB) function, and a sum of two blackbody functions with 
the normalization proportional to the surface area (2BB), which have been shown to be the best-fits to 
the broadband spectra of SGR bursts\cite{Feroci2004, Olive2004, Lin2012, Horst2012}.
The details of each model are presented below.

The power law model:
\begin{equation}
f_{\rmn{PL}} = A (E/E_n)^{\alpha} \mbox{ ;}
\end{equation}

the custom exponentially cutoff power law (CPL):
\begin{eqnarray*}
	n(E)& = &(E/E_n)^{\alpha} \exp(-E (2+\alpha)/E_\rmn{p}) \\
	f_{\rmn{CPL}} & = &F \times n(E) /\int_{E_\rmn{min}}^{E_\rmn{max}}n(E) E dE \mbox{ ;}
\end{eqnarray*}

the Band function:
\begin{equation*}
\label{eq:Band}
f_{\rmn{Band}} = A  \left\{ \begin{array}{ll}
(E/E_n)^{\alpha} \exp \left(-\frac{E (2+\alpha)}{E_\rmn{p}}\right), & \quad E<(\alpha-\beta)\frac{E_\rmn{p}}{2+\alpha} \\ 
(E/E_n)^{\beta} \left[\frac{E_\rmn{p} (\alpha-\beta)}{E_n (2+\alpha)}\right]^{(\alpha-\beta)} \exp(\beta-\alpha), & \quad E \ge(\alpha-\beta)\frac{E_\rmn{p}}{2+\alpha}, \\ 
\end{array} \right.
\end{equation*}
where  $f$ is a photon spectrum, measured in cm$^{-2}$~s$^{-1}$~keV$^{-1}$, $A$ is a model normalization, 
$E_n =100$~keV is a pivot energy, $E_\rmn{p}$ is the peak energy of the $\nu F_{\nu}$ spectrum, 
and $F$ is the model energy flux in the $E_\rmn{min}$ -- $E_\rmn{max}$ energy band;
$\alpha$ and $\beta$ are the low-energy and high-energy photon indices, respectively.
The single blackbody (BB) function and 2BB models are a single \texttt{bbodyrad} 
and a sum of two \texttt{bbodyrad} XSPEC models, respectively. The results of our spectral analysis are presented in Extended Data Tables 2 and 3.

The Poisson data with Gaussian background statistic (PG-stat) was used in the model fitting process as a figure of merit
to be minimized. The spectral channels were grouped to have a minimum of one counts per channel to ensure the validity of the fit statistic.

The best-fit spectral parameters and their confidence intervals were calculated using Bayesian statistics and a Markov chain Monte Carlo (MCMC) technique.   
We  performed  MCMC  simulations using  the XSPEC implementation of the Goodman-Weare algorithm 
where an ensemble of "walkers", which are vectors of the fit parameters, 
"explore" the parameter space via random steps determined by the walker positions on the previous iteration. 
We evolved eight walkers for a total of $4 \times 10^6$ steps, after discarding ("burn") the initial 10000 steps to ensure the chain reached a steady state. 
We use PG-stat likelihood and Gaussian priors centered on the initial fit result with variance matrix based 
on the covariance matrix of the fit. We then marginalized over the model normalization parameter  
to generate posterior probabilities for $\alpha$ and $E_\mathrm{p}$, using the XSPEC \texttt{margin} command. 
Then we find the best-fit parameters as a peak of the marginalized posterior distribution. 
The marginalized posterior distributions presented in Extended Data Fig.~2 in terms of integrated probability were produced directly from 
chain FITS files using Python3 programming language libraries Numpy\cite{Oliphant2006} and Matplotlib\cite{Hunter2007}.
in a way similar to the XSPEC \texttt{plot margin} command.
The contours encompassing the particular probability were constructed using Matplotlib \texttt{contour} function 
from the integrated probability distribution.

\paragraph*{Burst energetics}
The total energy fluence $S$ and the peak energy flux $F_\rmn{peak}$ of the burst
were derived using the energy flux of the best-fit CPL spectral model in the 20~keV--500~keV band.
Since the time-integrated spectrum accumulation interval differs from the $T_{100}$ interval
a correction which accounts for the emission outside the time-integrated spectrum
was introduced when calculating $S$.
$F_\rmn{peak}$ was calculated on the 16~ms scale using the best-fit CPL spectral
model for the spectrum near the peak count rate.
To obtain $F_\rmn{peak}$, the model energy flux was multiplied
by the ratio of the peak count rate to the average count rate in the spectral accumulation interval.
The Konus-Wind spectrum presented in Fig.~2 was measured over the interval from $T_0$ to $T_0+256$~ms 
and the derived flux was scaled to the average count rate in the interval $\pm 64$~ms around $T_0$,
which covers two main peaks of the burst. Both corrections mentioned above were made using counts in the G1+G2 light curve.

\subsection*{Emission scenarios}
To explain properties of simultaneous radio and X/$\gamma$-ray bursts from SGR 1935+2134 it is tempting to use approaches 
developed for FRBs~\cite{Cordes2019}, as magnetar scenarios for these extragalactic transients seem to be the most reliable, now.
Magnetar models of FRBs can be divided into two broad groups based on the regions where the emission is generated.
The first group of models~\cite{2014MNRAS.442L...9L,2017ApJ...842...34W,2018MNRAS.481.2407M}
assumes emission from a relativistic outflow which interacts with the surrounding medium. 
Derived from our analysis, the CPL photon index $\alpha$ has relatively wide confidential intervals extending to the values about -1.5, 
which are close to the indices typical for the fast cooling synchrotron regime widely discussed in GRB spectral models~\cite{1998ApJ...497L..17S, 2006RPPh...69.2259M,1999PhR...314..575P, 2015PhR...561....1K, 2019Galax...7...33P}. 
While the indices of $\alpha \leq - 1$ are less likely from the KW observations, 
they do allow to speculate about a single power law spectrum of the event extending from the radio to hard X-rays (see Fig.~2). 
The single power law can be expected in a relativistic outflow with internal shocks \cite{2017ApJ...843L..26B} 
or colliding magnetized shells \cite{2020arXiv200102007L} driven by the magnetospheric flares in the magnetar wind.     
In the later scenario the magnetic pulses produced by the magnetospheric flares of the total luminosity $L \geq 10^{41}$ erg/s 
propagate beyond the light cylinder (of radius 1.5$\times 10^{10}$~cm) with Lorentz factor $\Gamma \sim$15.  
They can collide at the dissipation region of the internal shells at the distance 
$\sim ct_{\rm var} \Gamma^2 \sim  3.4\times 10^{10}$~cm \cite{2006RPPh...69.2259M} 
i.e. just beyond the light cylinder for the flare parameters. 
The estimated value of magnetic field of the pulse in the dissipation region $\sim 10^4$~G. 
The electron-positrons in the colliding magnetized shells can be accelerated up 
to the Lorentz factor $\gamma \gtrsim 10^4$ over the time of 10$^{-5}$ s which is enough to produce 
the hard X-ray synchrotron photons with particle spectral indices $\geq$ 1 \cite{ABEtal2012, 2017SSRv..207..291B}.    
In an alternative scenario of magnetar transient emission~\cite{2002ApJ...580L..65L}, 
recently applied to FRBs~\cite{2020ApJ...889..135L}, radio/X-ray flares resemble solar flares.
Non-thermal particles are generated due to reconnection in a magnetar magnetosphere, 
i.e. below the light cylinder. Radio emission is generated early after the reconnection. 
If this radiation originates at a given distance from a neutron star, 
it is confined in a relatively narrow band, but a spreard in emission distances result in a wider spectrum. 
X-ray emission, which mainly appears when many electron-positron pairs are produced, 
is expected to be thermalized. This can be compatible with the observations 
if thermal fit for the high energy part of the spectrum is used.

\subsection*{Unusually-hard magnetar X-ray bursts}
In Figure~3 we show parameters of the April 28 flare as compared to 21 bursts from SGR~1935+2154 detected by Konus-\textit{Wind} (Ridnaia et al., In Preparation) and 41 bursts detected by \textit{Fermi}-GBM~\cite{Lin2020}.
With $E_p\sim85$~keV derived from the KW detection, the April 28 burst is a clear outlier in the SGR~1935+2154 population. 
We estimated the probability density function (PDF) of the peak energies for the combined KW+GBM burst sample  
using Gaussian kernels at observed $E_p$ values with 1$\sigma$ error bandwidth, 
and found the probability of the April~28 event $E_p$ to be drawn from the rest of SGR~1935+2154 bursts of $\sim 2 \times 10^{-10}$, 
thus highlighting the peculiar hardness of the April 28 event.

Five more bursts differ among all magnetar bursts detected by KW\cite{AptekarSGR,Mazets1627,GCN8851,GCN8858,GCN8863} (Ridnaia et al., In Preparation) 
with untypical, high $E_p$ values: two bursts from SGR~1900+14; the extremely-bright flare from SGR~1627-41; and two very bright bursts from SGR~1550-5418.   
We didn't find any radio observations simultaneous to the burst times, so none of these bursts can be excluded from candidates to FRBs. We note, that the hard burst detected on January 25, 2009 from SGR~1550-5418 had a similar to the April~28 burst, two-peaked light curve, with nearly the same separation between two peaks; 
and also radio pulsations were detected from this source $\sim$8 hours after the burst~\cite{Atel1913}.

%
%
%

\section*{Acknowledgements}
SP acknowledges support from the Program of development of M.V.~Lomonosov Moscow State University
(Leading Scientfic School "Physics of stars, relativistic objects, and galaxies").
The calculations were partially done on computers of the RAS JSCC and
St.~Petersburg department of the RAS JSCC and at the Tornado subsystem of the St.~Petersburg Polytechnic University supercomputing centre. 
The Konus-Wind experiment is supported by the Russian State Space Agency ROSCOSMOS.

\section*{Author contributions statement}

Konus-$Wind$ (KW) is a joint Russian-US gamma-ray burst experiment
with the Russian gamma-spectrometer Konus on board the NASA \textit{Wind} spacecraft.
A.R., D.S., and D.F. performed the Konus-\textit{Wind} data analysis.
A.R., D.S., D.F., A.B. and S.P. contributed to the discussion of 
the results in the context of KW magnetar observations, magnetar emission models and FRB-magnetar connection. 
A.L. contributed to the KW spectral analysis.
S.G., A.T., and M.U. contributed to the KW data reduction.
R.A., S.G., and D.F. contributed to the KW design and calibrations.
R.A. was the principal investigator of the KW experiment and T.C. was the co-PI from the American side.
A.R., D.S., and D.F. wrote the manuscript with the contributions of A.B. and S.P.
All co-authors provided comments on the final version of the manuscript.

\section*{ Additional information}
\footnotesize

\noindent {\bf Correspondence and requests for materials} should be addressed to \texttt{ridnaia@mail.ioffe.ru}. \\




\newpage

\begin{table*}
	\centering
	\caption*{{\bf Extended Data Table 1.} 
		Temporal parameters: the burst start time relative to $T_0$ (2nd column), 
		the burst end time relative to $T_0$ (3rd column), the total burst duration (4th column),
		$T_{50}$ and $T_{90}$  durations (5th and 6th columns, respectively)
		are given for specific energy band (1st column).}
	\label{tab:durations}
	\begin{tabular}{lccccc}
		\hline
		Band    & $T_\textrm{start}$  & $T_\textrm{stop}$ & $T_{100}$ & $T_{50}$  & $T_{90}$ \\[2pt]
		& (s)                 & (s)               & (s)      & (s)       & (s)     \\ 
		\hline
		G1    & -0.238 & 0.180 & 0.418 &  $0.078_{-0.011}^{+0.010}$ &$0.326_{-0.052}^{+0.032}$ \\[3pt]
    		G2    & -0.100 & 0.234 & 0.334 &  $0.074_{-0.022}^{+0.021}$ &$0.238_{-0.064}^{+0.053}$ \\[3pt]
                 G1+G2 & -0.220 & 0.244 & 0.464 &  $0.082_{-0.009}^{+0.013}$ &$0.310_{-0.034}^{+0.065}$ \\[3pt]
		
		\hline
	\end{tabular}
\end{table*}

\begin{table*}
	\centering
	\caption*{{\bf Extended Data Table 2.} 
		Spectral fit results with thermal models (BB and 2BB) for two spectra: 
		time-averaged spectrum (0.0--0.256 s) and the peak spectrum (0.0--0.064 s),
		measured near the peak count rate. The times are given relative to $T_0$. 
		The errors are given at 90\% CL. Model normalizations are proportional 
		to the surface area ${R_{km}^2}/{d_{10}^2}$, where $R_{km}$ is the source
		radius in km and $d_{10}$ is the distance to the source in units of 10~kpc.
	}
	\label{tab:spec1}
	\begin{tabular}{ccccccc}
		\hline
		Interval    & Model & kT$_1$ & Norm$_1$ & kT$_2$ & Norm$_2$ & PG-stat / dof \\[2pt]
		(s)      &       & (keV)  &  $({R_{1,km}^2}/{d_{10}^2})$        & (keV)  &  $({R_{2,km}^2}/{d_{10}^2})$         &              \\
		\hline\\
		0.0--0.064 & BB    & $19.5_{-1.5}^{+1.7}$      & $3.6_{-1.0}^{+1.3}$  & ...                   & ...                      & 45.2 / 34 \\[3pt]
		0.0--0.064 & 2BB   & $13.6_{-4.1}^{+3.4}$      & $9.3_{-4.2}^{+13.6}$  & $35.4_{-10.1}^{+21.8}$ & $0.16_{-0.02}^{+0.94}$ & 25.8 / 32 \\[3pt]
		\\
		0.0--0.256 & BB    & $19.5_{-1.4}^{+1.5}$      & $1.4_{-0.3}^{+0.4}$  & ...                   & ...                      & 82.6 / 48 \\[3pt]
		0.0--0.256 & 2BB   & $11.2_{-3.8}^{+3.3}$      & $6.3_{-3.2}^{+15.2}$  & $31.3_{-6.7}^{+12.2}$  & $0.14_{-0.11}^{+0.32}$ & 49.1 / 46 \\[3pt]
		\hline
	\end{tabular}
\end{table*}

\begin{table*}
	\centering
	\caption*{{\bf Extended Data Table 3.} 
		Spectral fit results with non-thermal models (PL and CPL) for two spectra: 
		time-averaged spectrum (0.0--0.256 s) and the peak spectrum (0.0--0.064 s),
		measured near the peak count rate. The times are given relative to $T_0$.
		The errors are given at 90\% CL. 
	}
	\label{tab:spec2}
	\begin{tabular}{cccccc}
		\hline
		Interval   & Model & $\alpha$                & E$_\textrm{p}$   & Flux (20--500 keV)                   & PG-stat / dof \\[2pt]
		  (s)      &       &                         & (keV)            & $10^{-6}$~erg~cm$^{-2}$~s$^{-1}$     &               \\
		\hline\\
		0.0--0.064 & PL    & $-2.02_{-0.12}^{+0.12}$ & ...              & $6.7_{-0.8}^{+0.9}$                  & 62.7 / 34 \\[3pt]
		0.0--0.064 & CPL   & $-0.42_{-0.58}^{+0.60}$ & $82_{-10}^{+14}$ & $5.8_{-0.6}^{+0.8}$                  & 25.4 / 33 \\[3pt]
		\\
		0.0--0.256 & PL    & $-2.02_{-0.09}^{+0.10}$ & ...              & $2.7_{-0.3}^{+0.3}$                  & 85.8 / 48 \\[3pt]
		0.0--0.256 & CPL   & $-0.72_{-0.46}^{+0.47}$ & $85_{-10}^{+15}$ & $2.4_{-0.2}^{+0.3}$                  & 45.6 / 47 \\[3pt]
		\hline
		
	\end{tabular}
\end{table*}

\newpage

\begin{table*}
	\centering
	\caption*{{\bf Extended Data Table 4.}
		Peak energies $E_p$ (CPL model) and energy fluences of hard magnetar bursts detected by KW.
		The errors are given at 68\% CL. The fluences are estimated in the 20--500 keV band.\\
		$^a$~Two hard bursts from SGR~1900+14 were also reported by BATSE\cite{WoodsHardBursts}.\\ 
		$^b$~The parameters were estimated using optically thin thermal bremsstrahlung (OTTB) spectrum ($f(E) \propto E^{-1}\exp(-E/kT_\mathrm{OTTB}))$.\\
		$^c$~The 2009-01-25 burst from SGR~1550-5418 was also reported by \textit{INTEGRAL}\cite{ATel1908,Mereghetti2009}.\\
		$^d$~The 2009-01-29 burst from SGR~1550-5418 was also detected by {\it{Fermi}}-GBM (https://gammaray.nsstc.nasa.gov/gbm/science/magnetars.html).\\
		$^e$~The FRB burst (this work).
	}
	\label{tab:kwhardEplist}
	
	\begin{tabular}{lclcc}
		\hline
		Date          & KW trigger time & Source 			& E$_\textrm{p}$   		& Fluence  					\\[3pt]
		&                 &                   & (keV)                 & $10^{-6}$~erg/cm$^{2}$   	\\
		\hline\\
		19981022\cite{AptekarSGR}$^{,a,b}$  &  15:40:46.627  & SGR~1900+14 	&  $110 \pm 30$  		&  $0.39_{-0.04}^{+0.04}$  	\\[3pt]
		19990110\cite{AptekarSGR}$^{,a,b}$  &  08:39:01.078  & SGR~1900+14 	&  $75 \pm 6$  	    	&  $0.80_{-0.06}^{+0.05}$  	\\[3pt]
		19980618\cite{Mazets1627}$^{,b}$  &  01:42:33.495  & SGR~1627-41  		&  $\sim100-150$  			&  $\sim700$  				\\[3pt]
		20090125\cite{GCN8851,GCN8858}$^{,c}$  &  08:43:37.781  & SGR~1550-5418  	&  $\sim70-100$  			&  $\sim250$  				\\[3pt]
		20090129\cite{GCN8863}$^{,d}$  &  09:17:07.224  & SGR~1550-5418  	&  $\sim90-100$  			&  $\sim50$  				\\[3pt]
		20200428$^e$  &  14:34:24.084  & SGR~1935+2154  	&  $85_{-6}^{+7}$  		&  $0.97_{-0.04}^{+0.04}$  	\\[3pt]
		
		\hline
		
	\end{tabular}
\end{table*}

\newpage
\begin{figure}\label{fig:spec}
\centering
\includegraphics[width=0.99\textwidth]{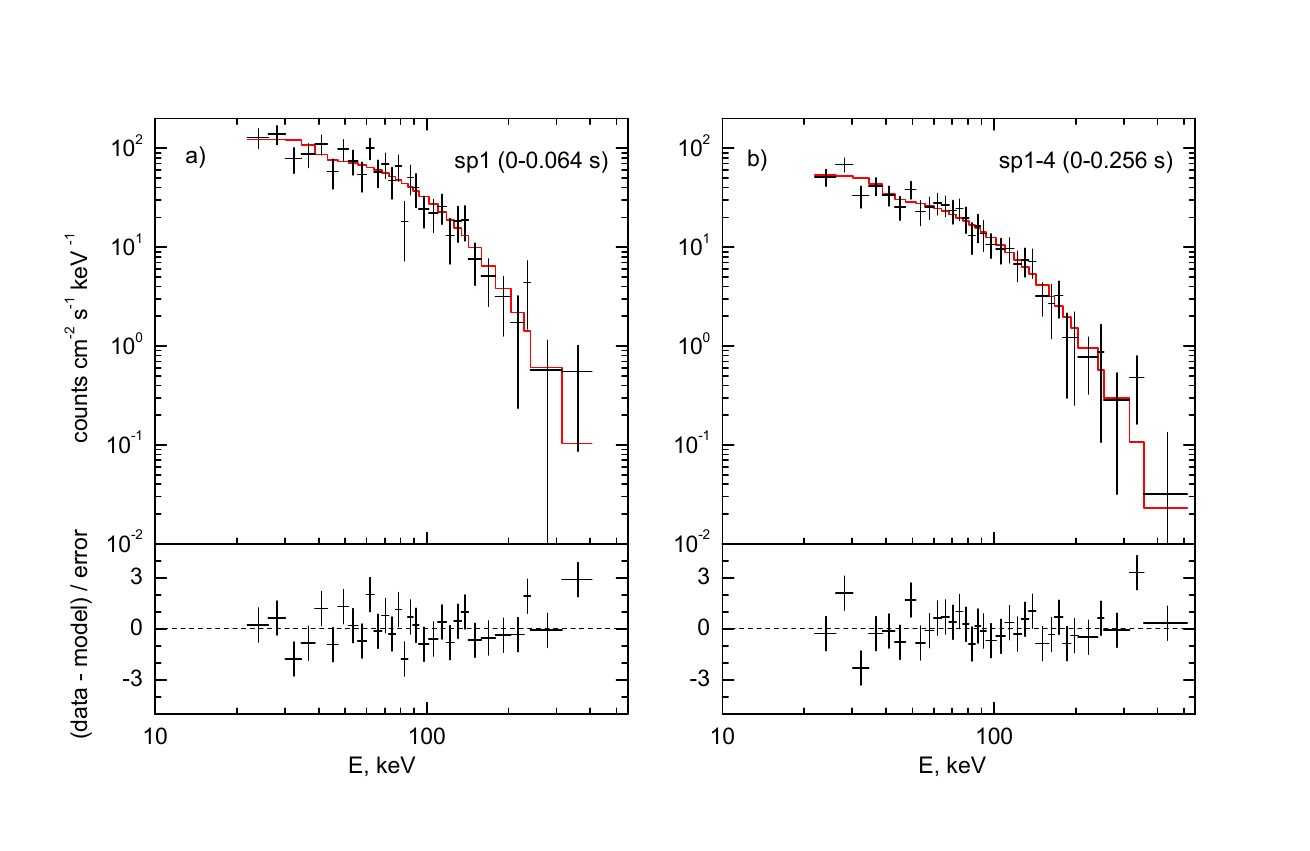}
\caption*{{\bf Extended Data Figure 1.} 
Konus-\textit{Wind} energy spectra of the April 28 event: 
(a) the spectrum containing peak count rate ($T_0$--$T_0$+0.064~s); (b) the time-integrated spectrum ($T_0$--$T_0$+0.256~s).
The top panels present the count rates and their uncertainties (black points) and the best-fit CPL model (red line). 
The bottom panels show the fit residuals.
Spectral channels are grouped for display purposes.}
\end{figure}

\begin{figure}\label{fig:KM}
	\centering
	\includegraphics[width=0.99\textwidth]{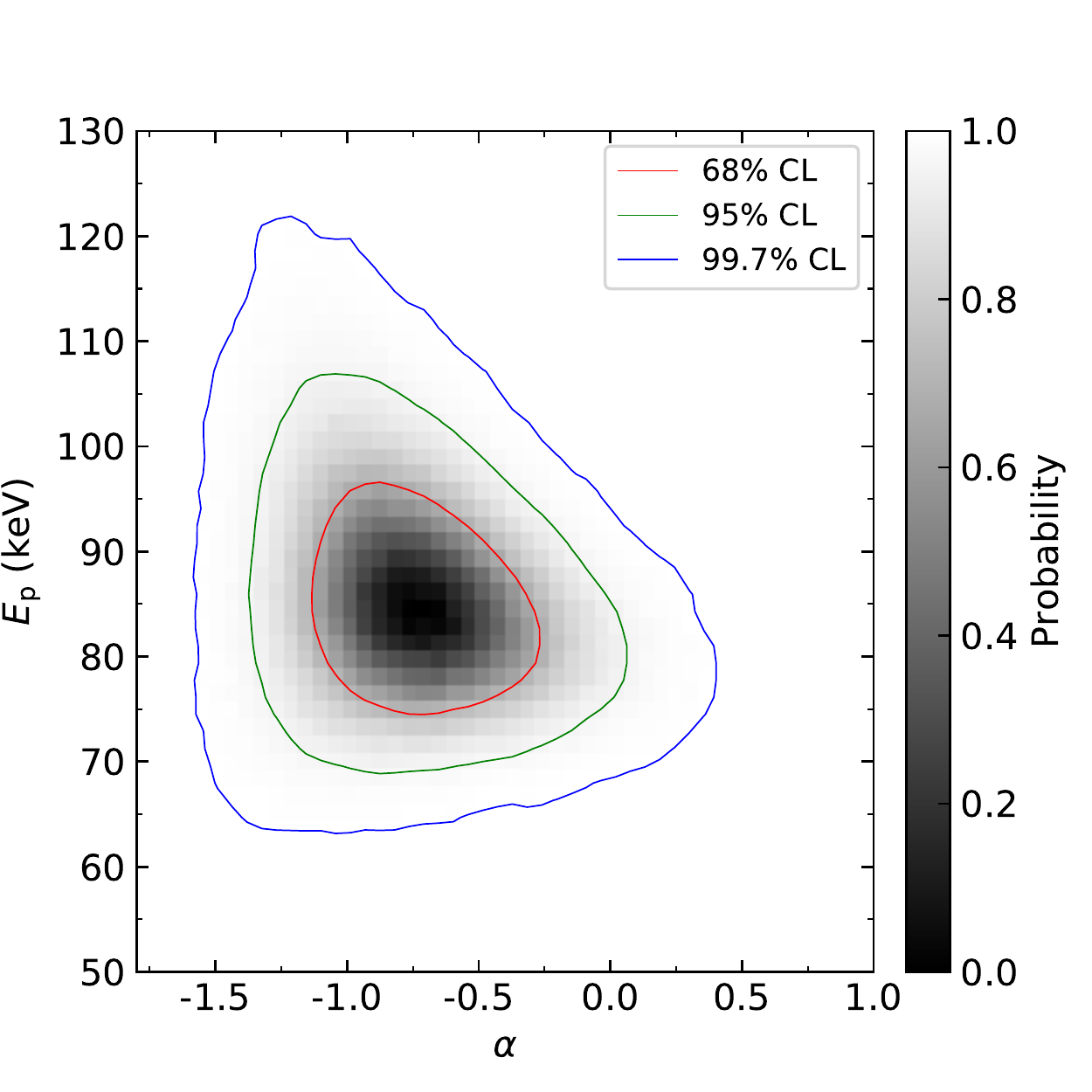}
	\caption*{{\bf Extended Data Figure 2.} 
	Integrated probability of posterior marginal distribution for CPL model parameters $\alpha$ and $E_p$ (the time-integrated spectrum). 
	The regions of the parameter space encompassing 68\%, 95\%, and 99.7\% of the distribution probability are shown with red, green, and blue contours, respectively.
}
\end{figure}

\begin{figure}\label{fig:KWmagnetars}
\centering
\includegraphics[width=0.8\textwidth]{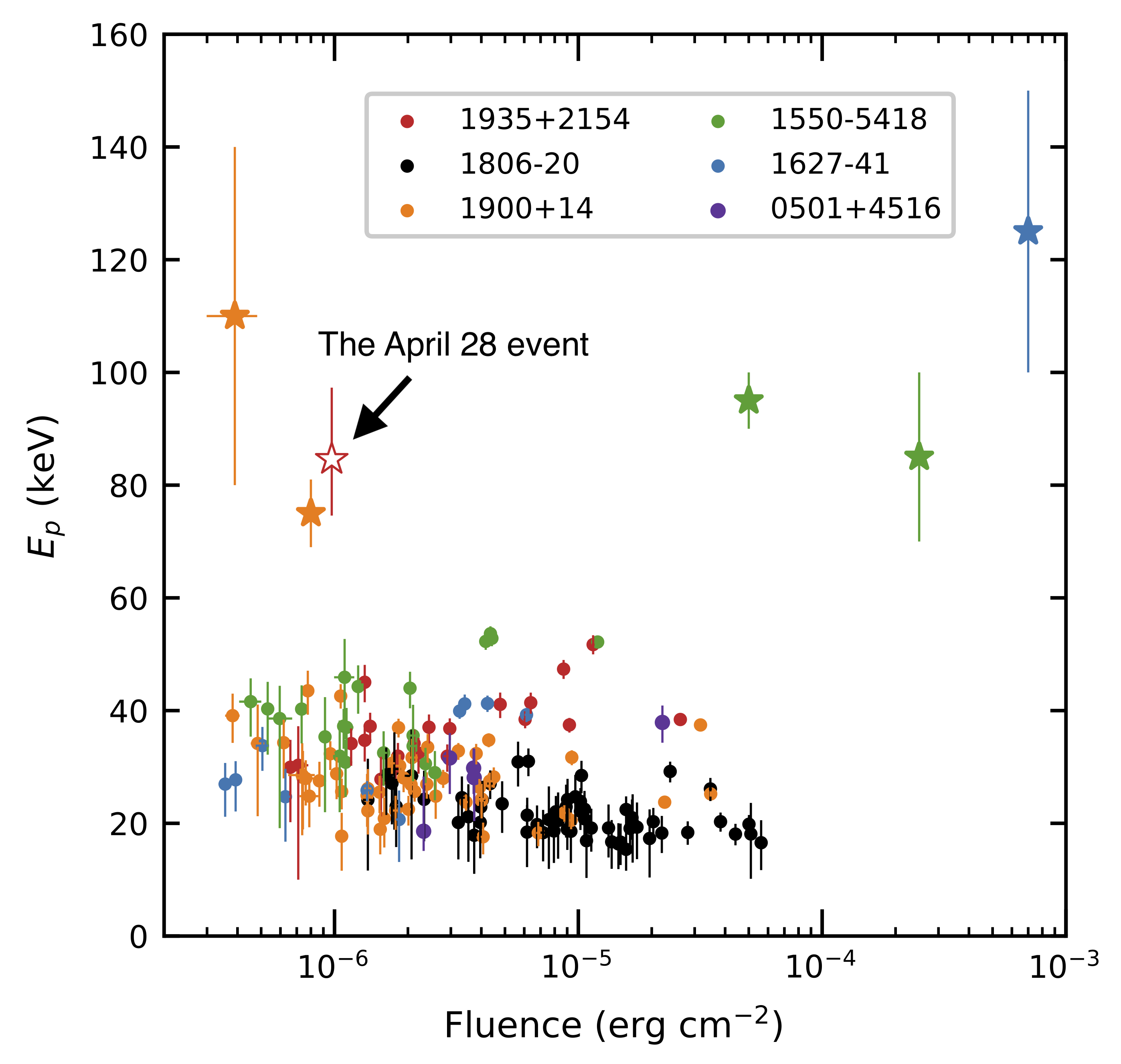}
\caption*{{\bf Extended Data Figure 3.} 
Peak energies $E_p$ vs. total energy fluences of $\sim$250 bright magnetar bursts detected by Konus-\textit{Wind} since November~1994. 
Six bursts marked by stars have unusually high $E_p$ values (Extended Data Table~4): 
two bursts from SGR~1900+14 (orange), one from SGR~1627-41 (blue), two from SGR~1550-5418 (green), and the April 28 event (red). 
The error bars are given at the 68\% confidence level. 
For extremely bright bursts from SGR~1627-41 and SGR~1550-5418 (shown by stars) the error bars represent systematic uncertainties due to the high count rate, which exceed statistical errors.
}

\end{figure}

\newpage
\section*{}
\newpage
\newcommand{\araa}{ARA\&A}   \newcommand{\aap}{Astron. Astrophys.}
\newcommand{\aj}{Astron. J.}         \newcommand{\apj}{Astrophys. J.}
\newcommand{\apjl}{Astrophys. J.}      \newcommand{\apjs}{Astrophys. J. Supp.}
\newcommand{\mnras}{Mon. Not. R. Astron. Soc.}   \newcommand{\nat}{Nature}
\newcommand{\pasj}{Publ. Astron. Soc. Japan}     \newcommand{\pasp}{Publ. Astron. Soc. Pac.}
\newcommand{\procspie}{Proc.\ SPIE} \newcommand{\physrep}{Phys. Rep.}
\newcommand{\apss}{APSS}
\newcommand{\solphys}{Sol. Phys.}
\newcommand{\actaa}{Acta Astronom}
\newcommand{\aaps}{Astron. Astrophys. Supp.}
\newcommand{\iaucirc}{IAU Circular}
\newcommand{\prd}{Phys. Rev. D}
\newcommand{\aapr}{Astron. Astrophys. Rev.}
\newcommand{\ssr}{Space Sci. Rev.}


\begin{thebibliography}{10}
\expandafter\ifx\csname url\endcsname\relax
  \def\url#1{\texttt{#1}}\fi
\expandafter\ifx\csname urlprefix\endcsname\relax\def\urlprefix{URL }\fi
\providecommand{\bibinfo}[2]{#2}
\providecommand{\eprint}[2][]{\url{#2}}

\bibitem{Cordes2019}
\bibinfo{author}{{Cordes}, J.~M.} \& \bibinfo{author}{{Chatterjee}, S.}
\newblock \bibinfo{title}{{Fast Radio Bursts: An Extragalactic Enigma}}.
\newblock \emph{\bibinfo{journal}{\araa}} \textbf{\bibinfo{volume}{57}},
\bibinfo{pages}{417--465} (\bibinfo{year}{2019}).
\newblock \eprint{1906.05878}.

\bibitem{Ravi2019}
\bibinfo{author}{{Ravi}, V.} \emph{et~al.}
\newblock \bibinfo{title}{{A fast radio burst localized to a massive galaxy}}.
\newblock \emph{\bibinfo{journal}{\nat}} \textbf{\bibinfo{volume}{572}},
  \bibinfo{pages}{352--354} (\bibinfo{year}{2019}).
  
  \bibitem{Marcote2020}
\bibinfo{author}{{Marcote}, B.} \emph{et~al.}
\newblock \bibinfo{title}{{A repeating fast radio burst source localized to a nearby spiral galaxy}}.
\newblock \emph{\bibinfo{journal}{\nat}} \textbf{\bibinfo{volume}{577}},
  \bibinfo{pages}{190--194} (\bibinfo{year}{2020}).
\newblock \eprint{2001.02222}.

\bibitem{CHIMErep}
\bibinfo{author}{{CHIME/FRB Collaboration}}
\newblock \bibinfo{title}{{CHIME/FRB Discovery of Eight New Repeating Fast Radio Burst Sources}}.
\newblock \emph{\bibinfo{journal}{\apjl}} \textbf{\bibinfo{volume}{885}},
  \bibinfo{pages}{L24} (\bibinfo{year}{2019}).
\newblock \eprint{1908.03507}.

\bibitem{2010vaoa.conf..129P}
\bibinfo{author}{{Popov}, S.~B.} \&  \bibinfo{author}{{Postnov}, K.~A.}
\newblock \bibinfo{title}{{Hyperflares of SGRs as an engine for millisecond extragalactic radio bursts}}.
\newblock \emph{\bibinfo{journal}{Evolution of Cosmic Objects through their Physical Activity}},
\bibinfo{pages}{129--132} (\bibinfo{year}{2010}).
\newblock \eprint{0710.2006}.

\bibitem{2017ApJ...843L..26B}
\bibinfo{author}{{Beloborodov}, A.~M.}
\newblock \bibinfo{title}{{A Flaring Magnetar in FRB 121102?}}.
\newblock \emph{\bibinfo{journal}{\apjl}} \textbf{\bibinfo{volume}{843}},
  \bibinfo{pages}{L26} (\bibinfo{year}{2017}).
\newblock \eprint{1702.08644}.

\bibitem{NatCHIME}
\bibinfo{author}{{The CHIME/FRB Collaboration}} 
\newblock \bibinfo{title}{{A bright millisecond-duration radio burst from a Galactic magnetar}}.
\newblock \emph{\bibinfo{journal}{\nat}} \textbf{\bibinfo{volume}{587}},
  \bibinfo{pages}{54--58} (\bibinfo{year}{2020}).

\bibitem{NatSTARE}
\bibinfo{author}{{Bochenek}, C.} \emph{et~al.} 
\newblock \bibinfo{title}{{A fast radio burst associated with a Galactic magnetar}}.
\newblock \emph{\bibinfo{journal}{\nat}} \textbf{\bibinfo{volume}{587}},
  \bibinfo{pages}{59--62} (\bibinfo{year}{2020}).

\bibitem{FASTUL}
\bibinfo{author}{{Lin}, L.} \emph{et~al.} 
\newblock \bibinfo{title}{{Stringent upper limits on pulsed radio emission during an active bursting phase of the Galactic magnetar SGRJ1935+2154}}.
Preprint at http://arXiv.org/abs/2005.11479 (\bibinfo{year}{2020}).

\bibitem{Kirsten2020}
\bibinfo{author}{{Kirsten}, F.} \emph{et~al.} 
\newblock \bibinfo{title}{{Detection of two bright radio bursts from magnetar SGR 1935 + 2154}}.
\newblock \emph{\bibinfo{journal}{Nat Astron}} (\bibinfo{year}{2020}).
 
 \bibitem{2015SSRv..191..315M}
 \bibinfo{author}{{Mereghetti}, S.}, \bibinfo{author}{{Pons}, J.} \& \bibinfo{author}{{Melatos}, A.},
 \newblock \bibinfo{title}{{Magnetars: Properties, Origin and Evolution}}.
 \newblock \emph{\bibinfo{journal}{\ssr}} \textbf{\bibinfo{volume}{191}},
 \bibinfo{pages}{315-338} (\bibinfo{year}{2015}).
 \newblock \eprint{1503.06313}.
 
 \bibitem{2017ARAA..55..261K}
 \bibinfo{author}{{Kaspi}, V.~M.} \& \bibinfo{author}{{Beloborodov}, A.~M.}
 \newblock \bibinfo{title}{{Magnetars}}.
 \newblock \emph{\bibinfo{journal}{\araa}} \textbf{\bibinfo{volume}{55}},
 \bibinfo{pages}{261-301} (\bibinfo{year}{2017}).
 \newblock \eprint{1703.00068}.
 
 \bibitem{1992ApJ...392L...9D}
 \bibinfo{author}{{Duncan}, R.~C.} \& \bibinfo{author}{{Thompson}, C.}
 \newblock \bibinfo{title}{{Formation of Very Strongly Magnetized Neutron Stars: Implications for Gamma-Ray Bursts}}.
 \newblock \emph{\bibinfo{journal}{\apjl}} \textbf{\bibinfo{volume}{392}},
 \bibinfo{pages}{L9} (\bibinfo{year}{1992}).
 
 \bibitem{1979Natur.282..587M}
 \bibinfo{author}{{Mazets}, E.~P.}, \bibinfo{author}{{Golentskii}, S.~V.},
 \bibinfo{author}{{Ilinskii}, V.~N.},\bibinfo{author}{{Aptekar}, R.~L.} \& \bibinfo{author}{{Guryan}, Iu.~A.}
 \newblock \bibinfo{title}{{Observations of a flaring X-ray pulsar in Dorado}}.
 \newblock \emph{\bibinfo{journal}{\nat}} \textbf{\bibinfo{volume}{282}},
 \bibinfo{pages}{587--589} (\bibinfo{year}{1979}).
 
 \bibitem{1979SvAL....5..343M}
 \bibinfo{author}{{Mazets}, E.~P.}, \bibinfo{author}{{Golentskii}, S.~V.} \& \bibinfo{author}{{Guryan}, Y.~A.}
 \newblock \bibinfo{title}{{Soft gamma-ray bursts from the source B1900+14}}.
 \newblock \emph{\bibinfo{journal}{Soviet Astronomy Letters}} \textbf{\bibinfo{volume}{5}},
 \bibinfo{pages}{343} (\bibinfo{year}{1979}).
 
 \bibitem{1998Natur.393..235K}
 \bibinfo{author}{{Kouveliotou}, C.}  \emph{et~al.}
 \newblock \bibinfo{title}{{An X-ray pulsar with a superstrong magnetic field in the
 		soft {\ensuremath{\gamma}}-ray repeater SGR1806 - 20}}.
 \newblock \emph{\bibinfo{journal}{\nat}} \textbf{\bibinfo{volume}{393}},
 \bibinfo{pages}{235--237} (\bibinfo{year}{1998}).
 
 \bibitem{Olausen2014}
\bibinfo{author}{{Olausen}, S.~A.} \& \bibinfo{author}{{Kaspi}, V.~M.}
\newblock \bibinfo{title}{{The McGill Magnetar Catalog}}.
\newblock \emph{\bibinfo{journal}{\apjs}} \textbf{\bibinfo{volume}{212}},
 \bibinfo{pages}{22} (\bibinfo{year}{2014}).
\newblock \eprint{1309.4167}.
 
\bibitem{Stamatikos2014}
\bibinfo{author}{{Stamatikos}, M.}, \bibinfo{author}{{Malesani}, D.},
\bibinfo{author}{{Page}, K.~L.}\& \bibinfo{author}{{Sakamoto}, T.}
\newblock \bibinfo{title}{{GRB 140705A: Swift detection of a short burst}}.
\newblock \emph{\bibinfo{journal}{GRB Coordinates Network, Circular Service, 
No.~16520, \#1 (2014)}} \textbf{\bibinfo{volume}{16520}}
 (\bibinfo{year}{2014}).
 
 \bibitem{Lien2014}
\bibinfo{author}{{Lien}, A.~Y.} \emph{et~al.}
\newblock \bibinfo{title}{{GRB 140705A: Swift-BAT refined analysis of a possible newly discovered SGR 1935+2154}}.
\newblock \emph{\bibinfo{journal}{GRB Coordinates Network, Circular Service, 
No.~16522, \#1 (2014)}} \textbf{\bibinfo{volume}{16522}}
 (\bibinfo{year}{2014}).

\bibitem{Israel2016}
\bibinfo{author}{{Israel}, G.~L.} \emph{et~al.}
\newblock \bibinfo{title}{{The discovery, monitoring and environment of SGR J1935+2154}}.
\newblock \emph{\bibinfo{journal}{\mnras}} \textbf{\bibinfo{volume}{457}},
  \bibinfo{pages}{3448--3456} (\bibinfo{year}{2016}).
\newblock \eprint{arXiv:astro-ph/1601.00347}.

\bibitem{Gaensler2014}
\bibinfo{author}{{Gaensler}, B.~M.}
\newblock \bibinfo{title}{{GRB 140705A / SGR 1935+2154: Probable association with supernova remnant G57.2+0.8}}.
\newblock \emph{\bibinfo{journal}{GRB Coordinates Network, Circular Service, 
No.~16533, \#1 (2014)}} \textbf{\bibinfo{volume}{16533}}
 (\bibinfo{year}{2014}).
          
 \bibitem{Zhou2020}
\bibinfo{author}{{Zhou}, P.} \emph{et~al.} 
\newblock \bibinfo{title}{{Revisiting the distance, environment and supernova properties of SNR G57.2+0.8 that hosts SGR 1935+2154}}.
\newblock \emph{\bibinfo{journal}{\apj}} \textbf{\bibinfo{volume}{905}},
  \bibinfo{pages}{99} (\bibinfo{year}{2020}).

\bibitem{Kothes2018}
\bibinfo{author}{{Kothes}, R.}, \bibinfo{author}{{Sun}, X.},
  \bibinfo{author}{{Gaensler}, B.} \& \bibinfo{author}{{Reich}, W.}
\newblock \bibinfo{title}{{A Radio Continuum and Polarization Study of SNR G57.2+0.8 Associated with Magnetar SGR 1935+2154}}.
\newblock \emph{\bibinfo{journal}{\apj}} \textbf{\bibinfo{volume}{852}},
  \bibinfo{pages}{54} (\bibinfo{year}{2018}).
\newblock \eprint{1711.11146}.

\bibitem{GCN27663}
\bibinfo{author}{{Ricciarini}, S} \emph{et~al.} 
\newblock \bibinfo{title}{{high bursting activity of SGR 1935+2154: CGBM observations}}.
\newblock \emph{\bibinfo{journal}{GRB Coordinates Network, Circular Service, 
No.~27663, \#1 (2020)}} \textbf{\bibinfo{volume}{27663}}
 (\bibinfo{year}{2020}).
 
 \bibitem{GCN27665}
\bibinfo{author}{{Palmer}, D.~M.}
\newblock \bibinfo{title}{{A Forest of Bursts from SGR 1935+2154}}.
\newblock \emph{\bibinfo{journal}{GRB Coordinates Network, Circular Service, 
No.~27665, \#1 (2020)}} \textbf{\bibinfo{volume}{27665}}
 (\bibinfo{year}{2020}).
 
 \bibitem{GCN27667}
\bibinfo{author}{{Ridnaia}, A} \emph{et~al.} 
\newblock \bibinfo{title}{{Konus-Wind observation of a very intense bursting activity of SGR 1935+2154}}.
\newblock \emph{\bibinfo{journal}{GRB Coordinates Network, Circular Service, 
No.~27667, \#1 (2020)}} \textbf{\bibinfo{volume}{27667}}
 (\bibinfo{year}{2020}).
 
  \bibitem{Mereghetti2020}
\bibinfo{author}{{Mereghetti}, S.} \emph{et~al.} 
\newblock \bibinfo{title}{{INTEGRAL discovery of a burst with associated radio emission from the magnetar SGR 1935+2154}}.
\newblock \emph{\bibinfo{journal}{\apjl}} \textbf{\bibinfo{volume}{898}},
\bibinfo{pages}{L29} (\bibinfo{year}{2020}).
 
\bibitem{NatAstrAGILE}
\bibinfo{author}{{Tavani}, M.}  \emph{et~al.} 
\newblock \bibinfo{title}{{An X-ray burst from a magnetar enlightening the mechanism of fast radio bursts}}.
\newblock \emph{\bibinfo{journal}{Nat Astron}} (\bibinfo{year}{2021}).
 
  \bibitem{GCNKWFRB}
\bibinfo{author}{{Ridnaia}, A} \emph{et~al.} 
\newblock \bibinfo{title}{{Konus-Wind observation of hard X-ray counterpart of the radio burst from SGR 1935+2154}}.
\newblock \emph{\bibinfo{journal}{GRB Coordinates Network, Circular Service, 
No.~27669, \#1 (2020)}} \textbf{\bibinfo{volume}{27669}}
 (\bibinfo{year}{2020}).
 
\bibitem{NatInsight}
\bibinfo{author}{{Li}, C.~K.}  \emph{et~al.} 
\newblock \bibinfo{title}{{HXMT identification of a non-thermal X-ray burst from SGR J1935+2154 and with FRB 200428}}.
\newblock \emph{\bibinfo{journal}{Nat Astron}} (\bibinfo{year}{2021}).

\bibitem{Law2017}
\bibinfo{author}{{Law}, C.~J.} \emph{et~al.} 
\newblock \bibinfo{title}{{A Multi-telescope Campaign on FRB 121102: Implications for the FRB Population}}.
\newblock \emph{\bibinfo{journal}{\apj}} \textbf{\bibinfo{volume}{850}},
\bibinfo{pages}{76} (\bibinfo{year}{2017}).

\bibitem{Gourdji2019}
\bibinfo{author}{{Gourdji}, K.} \emph{et~al.} 
\newblock \bibinfo{title}{{A Sample of Low-energy Bursts from FRB 121102}}.
\newblock \emph{\bibinfo{journal}{\apjl}} \textbf{\bibinfo{volume}{877}},
\bibinfo{pages}{L19} (\bibinfo{year}{2019}).

\bibitem{Kumar2020}
\bibinfo{author}{{Kumar},~P.}  \emph{et~al.} 
\newblock \bibinfo{title}{{Extremely band-limited repetition from a fast radio burst source}}.
\newblock \emph{\bibinfo{journal}{\mnras}} \textbf{\bibinfo{volume}{500}},
\bibinfo{pages}{2525--31} (\bibinfo{year}{2021}).
\newblock \eprint{2009.01214}.
 
\bibitem{2005Natur.434.1098H}
\bibinfo{author}{{Hurley}, K.}  \emph{et~al.}
\newblock \bibinfo{title}{{An exceptionally bright flare from SGR 1806-20 and the origins of short-duration {\ensuremath{\gamma}}-ray bursts}}.
\newblock \emph{\bibinfo{journal}{\nat}} \textbf{\bibinfo{volume}{434}},
\bibinfo{pages}{1098--1103} (\bibinfo{year}{2005}).

\bibitem{2007AstL...33....1F}
\bibinfo{author}{{Frederiks}, D.~D.}  \emph{et~al.}
\newblock \bibinfo{title}{{Giant flare in SGR 1806-20 and its Compton reflection from the Moon}}.
\newblock \emph{\bibinfo{journal}{Astronomy Letters}} \textbf{\bibinfo{volume}{33}},
\bibinfo{pages}{1--18} (\bibinfo{year}{2007}).
\newblock \eprint{ astro-ph/0612289}.

\bibitem{Svinkin2015}
\bibinfo{author}{{Svinkin}, D.~S.}, \bibinfo{author}{{Hurley}, K.}, \bibinfo{author}{{Aptekar}, R.~L.},
\bibinfo{author}{{Golenetskii}, S.~V.} \& \bibinfo{author}{{Frederiks}, D.~D.}
\newblock \bibinfo{title}{{A search for giant flares from soft gamma-ray repeaters in nearby galaxies in the Konus-WIND short burst sample}}.
\newblock \emph{\bibinfo{journal}{\mnras}} \textbf{\bibinfo{volume}{447}},
\bibinfo{pages}{1028--1032} (\bibinfo{year}{2015}).
\newblock \eprint{1411.5589}.

\bibitem{2013MNRAS.428.2857R}
\bibinfo{author}{{Rubio-Herrera}, E.}, \bibinfo{author}{{Stappers}, B.~W}, \bibinfo{author}{{Hessels}, J.~W.~T.}
\& \bibinfo{author}{{Braun}, R.} 
\newblock \bibinfo{title}{{A search for giant flares from soft gamma-ray repeaters in nearby galaxies in the Konus-WIND short burst sample}}.
\newblock \emph{\bibinfo{journal}{\mnras}} \textbf{\bibinfo{volume}{428}},
\bibinfo{pages}{2857--2873} (\bibinfo{year}{2013}).
\newblock \eprint{1210.4660}.

\bibitem{2008ApJ...680..545M}
\bibinfo{author}{{Mazets}, E.~P.} \emph{et~al.}
\newblock \bibinfo{title}{{A Giant Flare from a Soft Gamma Repeater in the Andromeda Galaxy (M31)}}.
\newblock \emph{\bibinfo{journal}{\apj}} \textbf{\bibinfo{volume}{680}},
\bibinfo{pages}{545--549} (\bibinfo{year}{2008}).
\newblock \eprint{0712.1502}.

\bibitem{2007AstL...33...19F}
\bibinfo{author}{{Frederiks}, D.~D.} \emph{et~al.}
\newblock \bibinfo{title}{{On the possibility of identifying the short hard burst GRB 051103 with a giant flare from a soft gamma repeater in the M81 group of galaxies}}.
\newblock \emph{\bibinfo{journal}{Astronomy Letters}} \textbf{\bibinfo{volume}{33}},
\bibinfo{pages}{19--24} (\bibinfo{year}{2007}).
\newblock \eprint{astro-ph/0609544}.

  \bibitem{GCN27595}
\bibinfo{author}{{Svinkin}, D.} \emph{et~al.} 
\newblock \bibinfo{title}{{Improved IPN error box for GRB 200415A (consistent with the Sculptor Galaxy)}}.
\newblock \emph{\bibinfo{journal}{GRB Coordinates Network, Circular Service, 
No.~27595, \#1 (2020)}} \textbf{\bibinfo{volume}{27595}}
 (\bibinfo{year}{2020}).
 
\bibitem{GCN27596}
\bibinfo{author}{{Frederiks}, D.} \emph{et~al.} 
\newblock \bibinfo{title}{{Konus-Wind observation of GRB 200415A (a magnetar Giant Flare in Sculptor Galaxy?))}}.
\newblock \emph{\bibinfo{journal}{GRB Coordinates Network, Circular Service, 
No.~27596, \#1 (2020)}} \textbf{\bibinfo{volume}{27596}}
 (\bibinfo{year}{2020}).
 
\bibitem{Kozlova2016}
\bibinfo{author}{{Kozlova}, A.~V.}  \emph{et~al.} 
\newblock \bibinfo{title}{{The first observation of an intermediate flare from SGR 1935+2154}}.
\newblock \emph{\bibinfo{journal}{\mnras}} \textbf{\bibinfo{volume}{460}},
\bibinfo{pages}{2008--2014} (\bibinfo{year}{2016}).
\newblock \eprint{1605.02993}.

\bibitem{Lin2020}
\bibinfo{author}{{Lin}, L.} \emph{et~al.}
\newblock \bibinfo{title}{{Burst Properties of the Most Recurring Transient Magnetar SGR J1935+2154}}.
\newblock \emph{\bibinfo{journal}{\apj}} \textbf{\bibinfo{volume}{893}},
\bibinfo{pages}{156} (\bibinfo{year}{2020}).
\newblock \eprint{2003.10582}.

\bibitem[44]{Aptekar_1995SSR}
\bibinfo{author}{{Aptekar}, R.} \emph{et~al.}
\newblock \bibinfo{title}{{Konus-W Gamma-Ray Burst Experiment for the GGS Wind Spacecraft}}.
\newblock \emph{\bibinfo{journal}{\ssr}} \textbf{\bibinfo{volume}{120}},
  \bibinfo{pages}{265--272} (\bibinfo{year}{1995}).

\bibitem[45]{Harten_1995SSRv}
\bibinfo{author} {{Harten}, R.} \& \bibinfo{author} {{Clark}, K.}
\newblock \bibinfo{title}{{The Design Features of the GGS Wind and Polar Spacecraft}}.
\newblock \emph{\bibinfo{journal}{\ssr}} \textbf{\bibinfo{volume}{71}},
  \bibinfo{pages}{23--40} (\bibinfo{year}{1995}).
  
\bibitem[46]{Mazets1999AstL}
\bibinfo{author}{{Mazets}, E.~P.} \emph{et~al.}
\newblock \bibinfo{title}{{Activity of the soft gamma repeater SGR 1900+14 in 1998 from Konus-Wind observations: 2.~The giant August 27 outburst}}.
\newblock \emph{\bibinfo{journal}{Astron.~Lett.}} 
  \textbf{\bibinfo{volume}{25}},  \bibinfo{pages}{635--648} 
  (\bibinfo{year}{1999}).
\newblock \eprint{arXiv:astro-ph/9905196}.

\bibitem[47]{Kouveliotou_1993ApJ}
\bibinfo{author}{{Kouveliotou}, C.} \emph{et~al.}
\newblock \bibinfo{title}{{Identification of two classes of gamma-ray bursts}}.
\newblock \emph{\bibinfo{journal}{\apjl}} \textbf{\bibinfo{volume}{413}},
  \bibinfo{pages}{L101--L104} (\bibinfo{year}{1993}).

\bibitem[48]{Koshut_1996}
\bibinfo{author}{{Koshut}, T.~M.} \emph{et~al.}
\newblock \bibinfo{title}{{Systematic Effects on Duration Measurements of Gamma-Ray Bursts}}.
\newblock \emph{\bibinfo{journal}{\apj}} \textbf{\bibinfo{volume}{463}},
  \bibinfo{pages}{570} (\bibinfo{year}{1996}).
  
\bibitem[49]{Arnaud1996}
\bibinfo{author}{{Arnaud}, K.~A.} \emph{et~al.}
\newblock \bibinfo{title}{{XSPEC: The First Ten Years}}.
\newblock \emph{\bibinfo{journal}{ASP~Conf.~Ser.}} 
\textbf{\bibinfo{volume}{101}},
  \bibinfo{pages}{17} (\bibinfo{year}{1996}).

\bibitem[50]{Band_1993ApJ}
\bibinfo{author}{{Band}, A.~J.} \emph{et~al.}
\newblock \bibinfo{title}{{BATSE observations of gamma-ray burst spectra. I - Spectral diversity}}.
\newblock \emph{\bibinfo{journal}{\apj}} 
\textbf{\bibinfo{volume}{413}},
\bibinfo{pages}{281--292} (\bibinfo{year}{1993}).
 
\bibitem[51]{Feroci2004}
\bibinfo{author}{{Feroci}, M.} \emph{et~al.}
\newblock \bibinfo{title}{{Broadband X-Ray Spectra of Short Bursts from SGR 1900+14}}.
\newblock \emph{\bibinfo{journal}{\apj}} 
\textbf{\bibinfo{volume}{612}},
  \bibinfo{pages}{408--413} (\bibinfo{year}{2004}).
\newblock \eprint{arXiv:astro-ph/0405104}.

\bibitem[52]{Olive2004}
\bibinfo{author}{{Olive}, J.-F.} \emph{et~al.}
\newblock \bibinfo{title}{{Time-resolved X-Ray Spectral Modeling of an Intermediate Burst from SGR 1900+14 Observed by HETE-2 FREGATE and WXM}}.
\newblock \emph{\bibinfo{journal}{\apj}} 
\textbf{\bibinfo{volume}{616}},
  \bibinfo{pages}{1148--1158} (\bibinfo{year}{2004}).
\newblock \eprint{arXiv:astro-ph/0403162}.

\bibitem[53]{Lin2012}
\bibinfo{author}{{Lin}, L.} \emph{et~al.}
\newblock \bibinfo{title}{{Broadband Spectral Investigations of SGR J1550-5418 Bursts}}.
\newblock \emph{\bibinfo{journal}{\apj}} 
\textbf{\bibinfo{volume}{756}},
  \bibinfo{pages}{54} (\bibinfo{year}{2012}).
\newblock \eprint{arXiv:1207.1434}.

\bibitem[54]{Horst2012}
\bibinfo{author}{{van der Horst}, A.~J.} \emph{et~al.}
\newblock \bibinfo{title}{{SGR J1550-5418 Bursts Detected with the Fermi Gamma-Ray Burst Monitor during its Most Prolific Activity}}.
\newblock \emph{\bibinfo{journal}{\apj}} 
\textbf{\bibinfo{volume}{749}},
  \bibinfo{pages}{122} (\bibinfo{year}{2012}).
\newblock \eprint{arXiv:1202.3157}.

\bibitem[55]{Oliphant2006}
\bibinfo{author}{{Oliphant}, T.~E.}
\newblock \bibinfo{title}{{A Guide to NumPy}}.
\newblock \emph{\bibinfo{journal}{Trelgol Publishing USA}}, 
(\bibinfo{year}{2006}).

\bibitem[56]{Hunter2007}
\bibinfo{author}{{Hunter}, J.~D.}
\newblock \bibinfo{title}{{Matplotlib: A 2D graphics environment}}.
\newblock \emph{\bibinfo{journal}{Comput.~Sci.~Eng.}} 
\textbf{\bibinfo{volume}{9}},
\bibinfo{pages}{90--95} (\bibinfo{year}{2007}).
 
\bibitem[57]{2014MNRAS.442L...9L}
\bibinfo{author}{{Lyubarsky}, Yu.}
\newblock \bibinfo{title}{{A model for fast extragalactic radio bursts.}}.
\newblock \emph{\bibinfo{journal}{\mnras}} \textbf{\bibinfo{volume}{442}},
\bibinfo{pages}{L9--L13} (\bibinfo{year}{2014}).
\newblock \eprint{1401.6674}.

\bibitem[58]{2017ApJ...842...34W}
\bibinfo{author}{{Waxman}, E.}
\newblock \bibinfo{title}{{On the Origin of Fast Radio Bursts (FRBs)}}.
\newblock \emph{\bibinfo{journal}{\apj}} \textbf{\bibinfo{volume}{842}},
\bibinfo{pages}{34} (\bibinfo{year}{2017}).
\newblock \eprint{1703.06723}.

\bibitem[59]{2018MNRAS.481.2407M}
\bibinfo{author}{{Margalit}, B.} \emph{et~al.} 
\newblock \bibinfo{title}{{Unveiling the engines of fast radio bursts, superluminous supernovae, and gamma-ray bursts}}.
\newblock \emph{\bibinfo{journal}{\mnras}} \textbf{\bibinfo{volume}{481}},
\bibinfo{pages}{2407--2426} (\bibinfo{year}{2018}).
\newblock \eprint{1806.05690}.

\bibitem[60]{1998ApJ...497L..17S}
\bibinfo{author}{{Sari}, R.}, \bibinfo{author}{{Piran}, T.} \&  \bibinfo{author}{{Narayan}, R}
\newblock \bibinfo{title}{{Spectra and Light Curves of Gamma-Ray Burst Afterglows}}.
\newblock \emph{\bibinfo{journal}{\apjl}} \textbf{\bibinfo{volume}{497}},
\bibinfo{pages}{L17--L20} (\bibinfo{year}{1998}).
\newblock \eprint{astro-ph/9712005}.

\bibitem[61]{1999PhR...314..575P}
\bibinfo{author}{{Piran}, T.} 
\newblock \bibinfo{title}{{Gamma-ray bursts and the fireball model}}.
\newblock \emph{\bibinfo{journal}{\physrep}} \textbf{\bibinfo{volume}{314}},
\bibinfo{pages}{575--667} (\bibinfo{year}{1999}).
\newblock \eprint{astro-ph/9810256}.

\bibitem[62]{2006RPPh...69.2259M}
\bibinfo{author}{{M{\'e}sz{\'a}ros}, P.} 
\newblock \bibinfo{title}{{Gamma-ray bursts}}.
\newblock \emph{\bibinfo{journal}{Rep. Prog. Phys.}},
\bibinfo{pages}{2259--2321} (\bibinfo{year}{2006}).
\newblock \eprint{astro-ph/0605208}.

\bibitem[63]{2015PhR...561....1K}
\bibinfo{author}{{Kumar}, P.} \& \bibinfo{author}{{Zhang}, B.}  
\newblock \bibinfo{title}{{The physics of gamma-ray bursts \&amp; relativistic jets}}.
\newblock \emph{\bibinfo{journal}{\physrep}} \textbf{\bibinfo{volume}{561}},
\bibinfo{pages}{1--109} (\bibinfo{year}{2015}).
\newblock \eprint{1410.0679}.

\bibitem[64]{2019Galax...7...33P}
\bibinfo{author}{{Pe'er}, A.} \& \bibinfo{author}{{Zhang}, B.}  
\newblock \bibinfo{title}{{Plasmas in Gamma-Ray Bursts: Particle Acceleration, 
Magnetic Fields, Radiative Processes and Environments}}.
\newblock \emph{\bibinfo{journal}{\physrep}} \textbf{\bibinfo{volume}{7}},
\bibinfo{pages}{33} (\bibinfo{year}{2019}).
\newblock \eprint{1902.02562}.

\bibitem[65]{2020arXiv200102007L}
\bibinfo{author}{{Lyubarsky}, Yu.} 
\newblock \bibinfo{title}{{Fast radio bursts from reconnection in magnetar magnetosphere}}.
\newblock \emph{\bibinfo{journal}{\apj}} \textbf{\bibinfo{volume}{897}},
\bibinfo{pages}{1} (\bibinfo{year}{2020}).
\newblock \eprint{2001.02007}.

\bibitem[66]{ABEtal2012}
\bibinfo{author}{{Bykov}, A.}  \emph{et~al.}
\newblock \bibinfo{title}{{Particle Acceleration in Relativistic Outflows}}.
\newblock \emph{\bibinfo{journal}{\ssr}} \textbf{\bibinfo{volume}{173}},
\bibinfo{pages}{309--339} (\bibinfo{year}{2012}).
\newblock \eprint{1205.2208}.

\bibitem[67]{2017SSRv..207..291B}
\bibinfo{author}{{Blandford}, R.}, \bibinfo{author}{{Yuan}, Y.}, \bibinfo{author}{{Hoshino}, M.} \&  \bibinfo{author}{{Sironi}, L.}
\newblock \bibinfo{title}{{Magnetoluminescence}}.
\newblock \emph{\bibinfo{journal}{\ssr}} \textbf{\bibinfo{volume}{207}},
\bibinfo{pages}{291--317} (\bibinfo{year}{2017}).
\newblock \eprint{1705.02021}.

\bibitem[68]{2002ApJ...580L..65L}
\bibinfo{author}{{Lyutikov}, M.} 
\newblock \bibinfo{title}{{Radio Emission from Magnetars}}.
\newblock \emph{\bibinfo{journal}{\apjl}} \textbf{\bibinfo{volume}{580}},
\bibinfo{pages}{L65--L68} (\bibinfo{year}{2002}).
\newblock \eprint{astro-ph/0206439}.

\bibitem[69]{2020ApJ...889..135L}
\bibinfo{author}{{Lyutikov}, M.} 
\newblock \bibinfo{title}{{Radius-to-frequency Mapping and FRB Frequency Drifts}}.
\newblock \emph{\bibinfo{journal}{\apj}} \textbf{\bibinfo{volume}{889}},
\bibinfo{pages}{135} (\bibinfo{year}{2020}).
\newblock \eprint{1909.10409}.

\bibitem[70]{AptekarSGR}
\bibinfo{author}{{Aptekar}, R.~L.}  \emph{et~al.}
\newblock \bibinfo{title}{{Konus catalog of SGR activity to 2000}}.
\newblock \emph{\bibinfo{journal}{\apjs}} \textbf{\bibinfo{volume}{137}},
\bibinfo{pages}{227} (\bibinfo{year}{2001}).

 \bibitem[71]{Mazets1627}
\bibinfo{author}{{Mazets}, E.~P.}  \emph{et~al.}
\newblock \bibinfo{title}{{Unusual Burst Emission from the New Soft Gamma Repeater SGR 1627-41}}.
\newblock \emph{\bibinfo{journal}{\apjl}} \textbf{\bibinfo{volume}{519}},
\bibinfo{pages}{L151-L153} (\bibinfo{year}{1999}).

\bibitem[72]{GCN8851}
\bibinfo{author}{{Golenetskii}, S.}  \emph{et~al.} 
\newblock \bibinfo{title}{{Konus-Wind detection of very bright SGR-like burst on January 25, 2009}}.
\newblock \emph{\bibinfo{journal}{GRB Coordinates Network, Circular Service, 
		No.~8851, \#1 (2009)}} \textbf{\bibinfo{volume}{8851}}
(\bibinfo{year}{2009}).

\bibitem[73]{GCN8858}
\bibinfo{author}{{Golenetskii}, S.}  \emph{et~al.} 
\newblock \bibinfo{title}{{Konus-Wind observation of AXP/SGR 1E1547.0-5408 bursting activity}}.
\newblock \emph{\bibinfo{journal}{GRB Coordinates Network, Circular Service, 
		No.~8851, \#1 (2009)}} \textbf{\bibinfo{volume}{8858}}
(\bibinfo{year}{2009}).

\bibitem[74]{GCN8863}
\bibinfo{author}{{Golenetskii}, S.}  \emph{et~al.} 
\newblock \bibinfo{title}{{Konus-Wind observation of bright burst from AXP/SGR 1E1547.0-5408 on January 29}}.
\newblock \emph{\bibinfo{journal}{GRB Coordinates Network, Circular Service, 
		No.~8863, \#1 (2009)}} \textbf{\bibinfo{volume}{8863}}
(\bibinfo{year}{2009}).

 \bibitem[75]{Atel1913}
\bibinfo{author}{{Burgay}, M.} \emph{et~al.}
\newblock \bibinfo{title}{{Back to radio: Parkes detection of radio pulses from the transient AXP 1E1547.0-5408}}.
\newblock \emph{\bibinfo{journal}{The Astronomer's Telegram, 
No.~1913, \#1 (2009)}} \textbf{\bibinfo{volume}{1913}}
 (\bibinfo{year}{2009}).
 
\bibitem[76]{WoodsHardBursts}
\bibinfo{author}{{Woods}, P.~M.}  \emph{et~al.}
\newblock \bibinfo{title}{{Hard Burst Emission from the Soft Gamma Repeater SGR 1900+14}}.
\newblock \emph{\bibinfo{journal}{\apjl}} \textbf{\bibinfo{volume}{527}},
\bibinfo{pages}{L47-L50} (\bibinfo{year}{1999}).
 
 \bibitem[77]{ATel1908}
\bibinfo{author}{{Baldovin}, C.} \emph{et~al.}
\newblock \bibinfo{title}{{INTEGRAL observes continued activity from AXP 1E1547.0-5408}}.
\newblock \emph{\bibinfo{journal}{The Astronomer's Telegram, 
No.~1908, \#1 (2009)}} \textbf{\bibinfo{volume}{1908}}
 (\bibinfo{year}{2009}).

\bibitem[78]{Mereghetti2009}
\bibinfo{author}{{Mereghetti}, S.}  \emph{et~al.}
\newblock \bibinfo{title}{{Strong Bursts from the Anomalous X-Ray Pulsar 1E~1547.0-5408 Observed with the INTEGRAL/SPI Anti-Coincidence Shield}}.
\newblock \emph{\bibinfo{journal}{\apjl}} \textbf{\bibinfo{volume}{649}},
\bibinfo{pages}{L74} (\bibinfo{year}{2009}).


\end{thebibliography}
\end{document}